\documentclass[preprint,showpacs,preprintnumbers,amsmath,amssymb]{revtex4}

\usepackage{graphicx}
\usepackage{revsymb}
\usepackage{dcolumn}
\usepackage{bm}

\begin{document}
\title{Time optimal control of a two-level dissipative quantum system}
\author{D. Sugny}
\email{dominique.sugny@u-bourgogne.fr}
\author{C. Kontz}
\author{H. R. Jauslin}
\affiliation{Institut Carnot, UMR CNRS 5027, BP 47870, 21078
Dijon, France}
\date{\today}
\begin{abstract}
We propose an analysis of the time-optimal control of a
dissipative two-level quantum system whose dynamics is governed by
the Lindblad equation. This simple system allows one to use tools
of geometric control theory and to construct its optimal synthesis
i.e. to determine the set of all the optimal trajectories starting
from a given initial point. We study different processes such as
conversion of a pure state into a mixed-state and purification of
a mixed-state. In particular cases, we show that dissipation is
not undesirable and can help accelerating the control.
\end{abstract}
 \maketitle
\section{Introduction}
Manipulating quantum system by using time-dependent electric field
remains a goal of primary interests in different physical
processes extending from the control of chemical reactions
\cite{warren,rabitz0,bifurcating} to quantum computing
\cite{nielsen}. In recent years, active research has been
performed to take into account the interaction of the system with
the environment which represents more realistic situations but
also more challenging control scenarios than for closed quantum
systems \cite{bacon,lloyd,solomon}. Different control strategies
have been proposed. In this paper we will not consider control
techniques using non-unitary control such as measurement (see for
instance the quantum Zeno effect)
\cite{gong,roa1,roa2,pechen,sugawara,shuang,mendes}, or other
strategies such as bang-bang pulses, strong coupling with another
system or control that can actively act on the dissipation (see
\cite{facchi} for a recent review and references therein). It has
been shown that such methods are particularly efficient and can
even halt decoherence. We restrict our study to unitary control.
In this case and for the interaction with a markovian or a
non-markovian bath, the fact that the control field cannot fully
compensate the effect of dissipation largely enhances the
difficulty of the control. This point has been rigourously shown
in Refs. \cite{altafini1,altafini2} for a dynamics governed by the
Lindblad equation \cite{lindblad,gorini}. In this context, several
 studies using numerical optimization techniques have
proved that efficient control can still be achieved
\cite{rabitz1,rabitz2,tannor,potz1,potz2,sugny1,sugny2}. Due to
the complexity of realistic systems with multiple degrees of
freedom, this purely numerical approach seems to be the only
possible way to achieve control. More geometrical aspects of the
control using mainly tools of geometric control theory
\cite{jurdjevic,boscain,bonnard} can be formulated only for
simplest quantum systems having few levels (typically two or
three) or consisting in the coupling of spin $1/2$ particles. This
has been done recently by a large number of mathematical papers
dealing with closed
\cite{boscain1,boscain2,boscain3,boscain4,khaneja1,khaneja2,khaneja3,sugny3,sugny4,sugny5}
 or dissipative \cite{khaneja4} quantum systems. One of the objectives of the control has been the minimization of
the total time of the process either with constraints
\cite{boscain1,boscain4} or no constraints on the laser intensity
\cite{khaneja1}.

In this paper, we propose to do a step towards the geometrical
analysis of the control of dissipative quantum systems by
beginning with the simplest system possible, a two-level system
governed by the Lindblad equation. We determine control fields
which minimize the total time of the process by applying the
Pontryagin maximum principle (PMP)
\cite{pontryagin,jurdjevic,boscain,bonnard}. The maximum of the
laser intensity is fixed to an arbitrary value. An increase of
this value leads to a reduction of the duration of the control.
Note also that this cost functional seems particularly relevant in
the context of a dissipative environment especially when the
effect of dissipation is undesirable to reach the objective of the
control. A second argument explaining our choice of cost
functional is of mathematical nature. The resolution of the PMP is
particularly simple in this case and can be done analytically. The
geometrical description of the time-optimal control is also well
developed especially on $\mathbb{R}^2$ (see \cite{boscain} for a
recent overview) for affine systems with a drift term
corresponding here to the dissipation. The main tools of this
construction will be recalled throughout the paper. Finally, all
this mathematical arsenal allows us to answer some physical
questions such as the benefit that can be gained from dissipation.

 The paper is organized as follows. We first introduce the model
for the Lindblad equation of a two-level system. Writing the
density matrix in the coherence vector form \cite{schrimer2} and
assuming some constraints on the control term, we restrict the
problem to a control on a closed submanifold of $\mathbb{R}^2$
with a true drift term i.e. an uncontrollable term which cannot be
eliminated by a unitary transformation. This drift term is due to
the dissipation. We consider different cases with different
initial and target states that represent several situations of
physical interest. We formulate in Sec. \ref{sec2} the PMP with a
time minimum cost functional. In Sec. \ref{sec3}, we analyze the
structure of the reachable set from each initial state and we
point out the particular role of the fixed point of the
free-dynamics in this structure. Section \ref{sec4} deals with the
construction of the optimal syntheses by solving the PMP. We
recall that the PMP allows us to derive extremal controls. Optimal
trajectories are a subset of this set which can be determined in a
second step by a direct comparison of different extremal
trajectories or by geometrical arguments. Conclusions and
prospective views are given in Sec.\ref{sec6}. Some technical
calculations are reported in appendices \ref{appa} and \ref{appb}.
\section{The model system} \label{sec1}
We consider a dissipative two-level quantum system whose dynamics
is governed by the Lindblad equation. The system is described by a
density operator $\rho(t)$ which is a positive semi-definite
hermitian operator acting on a two-dimensional Hilbert space
$\mathcal{H}$. The evolution equation can be written as
\begin{equation} \label{eq0}
i\frac{\partial \rho}{\partial
t}=[H_0+uH_1,\rho]+i\mathcal{L}_D(\rho) \ ,
\end{equation}
where $H_0$ is the field-free Hamiltonian of the system, $H_1$
represents the interaction with the control field and
$\mathcal{L}_D$ the dissipative part of the equation. In the
Lindblad equation, $\mathcal{L}_D$ can be written in the general
form
\begin{equation} \label{eq00}
\sum_{k=1}^3[L_k\rho L_k\dag-\frac{1}{2}(\rho L_k\dag L_k+L_k\dag
L_k\rho)] \ ,
\end{equation}
where the operators $L_k$ are given by
\begin{eqnarray} \label{eq001}
L_1= \sqrt{\gamma_{21}}\left(
\begin{array}{cc}
0 & 1 \\ 0 & 0
\end{array} \right);
L_2= \sqrt{\gamma_{12}}\left(
\begin{array}{cc}
0 & 0 \\ 1 & 0
\end{array} \right);
L_3= \sqrt{\tilde{\Gamma}}\left(
\begin{array}{cc}
1 & 0 \\ 0 & -1
\end{array} \right) \ .
\end{eqnarray}
$\gamma_{12}$, $\gamma_{21}$ and $\tilde{\Gamma}$ are positive
real constants describing the interaction with the environment.
$\gamma_{12}$ and $\gamma_{21}$ correspond to population
relaxations whereas $\tilde{\Gamma}$ is the pure dephasing rate.
We consider resonant fields, i.e. the frequency $\omega$ of the
laser is equal to the energy difference between the two levels. In
the RWA approximation, the time evolution of $\rho(t)$ satisfies
the following Redfield form of the Lindblad equation
\begin{eqnarray} \label{eq1}
i \frac{\partial}{\partial t} \left(
\begin{array}{c}
\rho_{11} \\ \rho_{12} \\ \rho_{21} \\ \rho_{22}
\end{array} \right) =
\left(
\begin{array}{cccc}
-i\gamma_{12} & -u^*e^{-i\omega t} & ue^{i\omega t} & i\gamma_{21} \\
-ue^{i\omega t} & -\omega-i\Gamma & 0 &  ue^{i\omega t} \\
u^*e^{-i\omega t} & 0 & \omega-i\Gamma & -u^*e^{-i\omega t} \\
i\gamma_{12} & u^*e^{-i\omega t} & -ue^{i\omega t} & -i\gamma_{21}
\end{array} \right)
\left(
\begin{array}{c}
\rho_{11} \\ \rho_{12} \\ \rho_{21} \\ \rho_{22}
\end{array} \right)
\ ,
\end{eqnarray}
where $u$ is the complex Rabi frequency of the laser field (the
real and imaginary parts are the amplitudes of two orthogonal
linearly polarized fields). $\Gamma$ is the total dephasing rate
which can be written as
\begin{equation} \label{eq2}
\Gamma=\frac{1}{2}(\gamma_{12}+\gamma_{21})+\tilde{\Gamma} \ .
\end{equation}
From Eqs. (\ref{eq2}), we also notice that the requirement that
Eq. (\ref{eq1}) corresponds to a Lindblad equation implies the
constraint $\Gamma\geq \frac{1}{2}(\gamma_{12}+\gamma_{21})$ or
equivalently
 $\tilde{\Gamma}\geq 0$. Equation (\ref{eq1}) is written in
units such that $\hbar=1$. In the interaction representation, Eq.
(\ref{eq1}) becomes
\begin{eqnarray} \label{eq3}
i \frac{\partial}{\partial t} \left(
\begin{array}{c}
\tilde{\rho}_{11} \\ \tilde{\rho}_{12} \\ \tilde{\rho}_{21} \\
\tilde{\rho}_{22}
\end{array} \right) =
\left(
\begin{array}{cccc}
-i\gamma_{12} & -u^* & u & i\gamma_{21} \\
-u & -i\Gamma & 0 &  u \\
u^* & 0 & -i\Gamma & -u^* \\
i\gamma_{12} & u^* & -u & -i\gamma_{21}
\end{array} \right)
\left(
\begin{array}{c}
\tilde{\rho}_{11} \\ \tilde{\rho}_{12} \\ \tilde{\rho}_{21} \\
\tilde{\rho}_{22}
\end{array} \right)
\ .
\end{eqnarray}
Since $\textrm{Tr}[\rho]=1$, the density matrix $\rho$ depends on
three real parameters which can be given by the coordinates of the
coherence vector \cite{schrimer2} : $x_1=2\Re[\tilde{\rho}_{12}]$,
$x_2=2\Im[\tilde{\rho}_{12}]$ and
$x_3=\tilde{\rho}_{22}-\tilde{\rho}_{11}$. From Eq. (\ref{eq3}),
one deduces that the coordinates $x_i$ satisfy the following
system of inhomogeneous linear differential equations
\begin{eqnarray}\label{eq4}
\left\{ \begin{array}{lll}
\dot{x_1}=-\Gamma x_1+u_2x_3 \\
\dot{x_2}=-\Gamma x_2-u_1x_3 \\
\dot{x_3}=(\gamma_{12}-\gamma_{21})-(\gamma_{12}+\gamma_{21})x_3+u_1x_2-u_2x_1
\end{array} \right. \ ,
\end{eqnarray}
$u_1$ and $u_2$ being two real functions such that $u=u_1+iu_2$.
As $\textrm{Tr}[\rho^2]\leq 1$, we also have
$x_1^2+x_2^2+x_3^2\leq 1$  which defines the Bloch ball. The
dynamics is called either unital if $\gamma_{12}=\gamma_{21}$ i.e.
the fixed point of the free dynamics is the center of the Bloch
ball or affine otherwise \cite{altafini1,altafini2}.

To simplify the study, we restrict the dynamics to a submanifold
$M\subset\mathbb{R}^2$ by assuming that the control field is real
i.e. $u_2=0$. This hypothesis means that the control field is
linearly polarized. With this choice of control, the last two
equations of the system of Eqs. (\ref{eq4}) are decoupled from the
first one and the problem is reduced to $\mathbb{R}^2$. The system
of Eqs. (\ref{eq4}) then becomes
\begin{eqnarray}\label{eq5}
\left\{ \begin{array}{ll}
\dot{x_2}=-\Gamma x_2-ux_3 \\
\dot{x_3}=\gamma_--\gamma_+x_3+ux_2
\end{array} \right. \ ,
\end{eqnarray}
where $\gamma_-=\gamma_{12}-\gamma_{21}$ and
$\gamma_+=\gamma_{12}+\gamma_{21}$. The coordinate $x_1$ is set
initially to 0. To simplify the notation, the index 1 of $u_1$ has
been omitted when confusion is unlikely to occur. Note that the
analysis of the optimal control on $\mathbb{R}^3$ is considerably
more complex \cite{bonnard} and goes beyond the scope of this
paper. Equations (\ref{eq5}) can be written in a more compact form
\begin{equation} \label{eq6}
\dot{\textbf{x}}=F+uG \ ,
\end{equation}
with the vector $\textbf{x}$ of coordinates $(x_2,x_3)$ and the
two vector fields $F$ and $G$ defined by
\begin{eqnarray} \label{eq7}
F=\left(
\begin{array}{c}
-\Gamma x_2 \\ \gamma_--\gamma_+x_3
\end{array} \right) \textrm{and} \ G=
\left(
\begin{array}{c}
-x_3 \\ x_2
\end{array} \right)
\ .
\end{eqnarray}
Finally, a straightforward calculation shows that the fixed-point
of the free-dynamics is given for $\gamma_+\neq 0$ by
\begin{eqnarray}\label{eq8}
\left\{ \begin{array}{ll}
x_2=0 \\
x_3=\frac{\gamma_-}{\gamma_+}
\end{array} \right. \ ,
\end{eqnarray}
which is therefore either affine or unital according to the value
of $\gamma_-$. The irreversibility of the dissipation effects is
reflected in the fact that the vector field $F$ is a true drift
term which cannot be eliminated by unitary transformations
\cite{boscain1}.
\section{Pontryagin maximum principle} \label{sec2}
We analyze the optimal control of this two-level system with the
constraint of minimizing the total time of the control. We assume
that the field $u$ is bounded by
\begin{equation} \label{eq9}
|u|\leq 1 \ .
\end{equation}
Equations (\ref{eq5}) being linear, other bounds for $u$ can be
considered from a standard rescaling of the time and the
dissipative constants $\Gamma$, $\gamma_+$ and $\gamma_-$. The
Pontryagin maximum principle
\cite{pontryagin,jurdjevic,boscain,bonnard} is formulated from the
following pseudo-Hamiltonian $H_P$
\begin{equation} \label{eq10}
H_P=\textbf{p}\cdot(F+uG)+p_0 \ ,
\end{equation}
where $\textbf{p}=(p_2,p_3)\in (\mathbb{R}^2)^*$ is called the
adjoint state and $p_0$ is a negative constant. We recall that the
cost function for the minimum time problem is equal to 1. This
term is multiplied by the constant $p_0$ in the Hamiltonian $H_P$.
The Pontryagin maximum principle states that the extremal
trajectories maximize $H_P$ i.e.
\begin{equation} \label{eq11}
H_{max}(\textbf{x},\textbf{p})=Max_{|u|\leq
1}H_P(\textbf{x},\textbf{p},u) \ .
\end{equation}
The coordinates of the extremal vector state $\textbf{x}$ and of
the corresponding adjoint state $\textbf{p}$ fulfill the Hamilton
equations
\begin{eqnarray}\label{eq11a}
\left\{ \begin{array}{ll}
\dot{x_2}=-\Gamma x_2-ux_3 \\
\dot{x_3}=\gamma_--\gamma_+x_3+ux_2
\end{array} \right. \ ,
\end{eqnarray}
and
\begin{eqnarray}\label{eq11b}
\left\{ \begin{array}{ll}
\dot{p_2}=\Gamma p_2-up_3 \\
\dot{p_3}=\gamma_+p_3+up_2
\end{array} \right. \ ,
\end{eqnarray}
where $u$ is here the extremal control given by Eq. (\ref{eq11}).\\

We can now pass to the construction of the optimal syntheses. The
construction begins with the introduction of two sets of points
$\Delta_A^{-1}(0)$ and $\Delta_B^{-1}(0)$ denoted $C_A$ and $C_B$
which divide $M$ in different regions \cite{boscain}. $\Delta_A$
and $\Delta_B$ are two functions from $M$ to $\mathbb{R}$ defined
as follows
\begin{eqnarray}\label{eq12}
\left\{ \begin{array}{ll}
\Delta_A(\textbf{x})=Det(F,G) \\
\Delta_B(\textbf{x})=Det(G,[F,G])
\end{array} \right. \ ,
\end{eqnarray}
where $Det$ is the determinant of two vector fields and $[.,.]$
their commutator. In our case, simple algebra leads to
\begin{eqnarray}\label{eq13}
\left\{ \begin{array}{ll}
\Delta_A(\textbf{x})=-\Gamma x_2^2+\gamma_- x_3-\gamma_+ x_3^2 \\
\Delta_B(\textbf{x})=2\Gamma x_2x_3-2\gamma_+ x_2x_3+\gamma_- x_2
\end{array} \right. \ .
\end{eqnarray}
The major role of these two sets in the resolution of the optimal
control problem will be detailed in Sec. \ref{sec4}. We can
already say that $C_A$ and $C_B$ are responsible for qualitative
modifications of the optimal trajectories \cite{boscain,bonnard}.
A preliminary step thus consists in analyzing the structure of the
sets $C_A$ and $C_B$ when $\Gamma$, $\gamma_+$ and $\gamma_-$ vary
 with the conditions $\gamma_+\geq 0$ and $\Gamma\geq \frac{\gamma_+}{2}$.
 From a formal point of view, this can be done by introducing the
feed-back group \cite{bonnard} but the dynamics being here
bilinear [Eqs. (\ref{eq5})], we can directly compute the sets
$C_A$ and $C_B$ and their relative positions.

\begin{figure}
\includegraphics[width=0.4\textwidth]{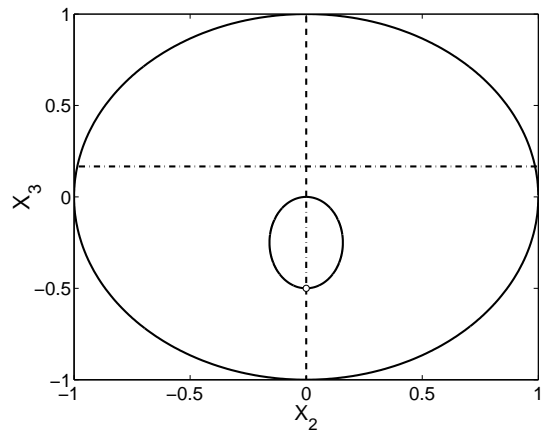}
\caption{\label{fig1} Division of the manifold $M$ by the sets
$C_A=\Delta_A^{-1}(0)$ (in solid line) and $C_B=\Delta_B^{-1}(0)$
(in dashed and dot-dashed lines) for $\gamma_-\neq 0$. The dashed
and dot-dashed lines represent respectively anti-turnpike and
turnpike arcs which are introduced in Sec. \ref{sec4}. The
exterior circle in solid line corresponds to the limit of the
Bloch ball in the plane $(x_2,x_3)$. The small open circle
indicates the position of the fixed-point of the free-dynamics.
Numerical values are taken to be $\Gamma=1$, $\gamma_{12}=0.1$ and
$\gamma_{21}=0.3$.}
\end{figure}

If $\Gamma\neq\gamma_+$ then the set $C_B$ corresponds to the
union of the two lines
\begin{equation}\label{eq14a}
x_2=0 \ ,
\end{equation}
and
\begin{equation}\label{eq14b}
x_3=\frac{-\gamma_-}{2\Gamma-2\gamma_+} \ ,
\end{equation}
with the restriction that $|x_3|\leq 1$. In the case
$\Gamma=\gamma_+$, $C_B$ is only composed of the vertical line of
equation $x_2=0$. The points $(x_2,x_3)$ of the Bloch ball that
are solutions of the polynomial equation
$\gamma_+x_3^2-\gamma_-x_3-\Gamma x_2^2=0$ belong to $C_A$. In the
case $\gamma_-\neq 0$, $C_A$ is therefore the union  of two
 parabolas. This set is either above or below the line
 $x_3=\frac{-\gamma_-}{2\Gamma-2\gamma_+}$ according to
the signs of $\gamma_-$ and of $\Gamma-\gamma_+$. For
$\gamma_-=0$, this set is reduced to the origin of the Bloch ball.
Figure \ref{fig1} displays the submanifold $M$ together with the
sets $C_A$ and $C_B$ for a given value of the parameters. We
finally notice that the fixed-point of the free dynamics and the
center of the Bloch ball belong to $C_A$.
\section{Reachable sets and controllability} \label{sec3}
We consider four different qualitative cases of control which
allow one to study several physically relevant situations :
\begin{itemize}
\item Case (a) : Conversion of a pure state into a mixed state with a unital
Lindbladian ($\gamma_-/\gamma_+=0$, $\Gamma>\gamma_+ +2$).
\item Case (b) : Conversion of a pure state into a mixed state with a unital
Lindbladian ($\gamma_-/\gamma_+=0$, $\gamma_+-2<\Gamma<\gamma_+
+2$).
\item Case (c) : Purification of the completely random mixed state which
corresponds to the center of the Bloch ball
($\gamma_-/\gamma_+=-1$, $\Gamma>\gamma_++2$).
\item Case (d) : Conversion of a pure state into a mixed state with an affine
Lindbladian ($\gamma_-/\gamma_+=-0.5$, $\Gamma>\gamma_++2$).
\end{itemize}
The numerical values we have chosen for illustrations are given in
Table \ref{tab1}.
\begin{table}[ht]
  \caption{\label{tab1} Numerical values of the dissipative constants in arbitrary units.}
  \begin{center}
\begin{tabular}{c|c|c|c}
\hline
 & $\Gamma$ & $\gamma_{12}$ &  $\gamma_{21}$ \\
\hline \hline
(a) & 3 & 0.3 & 0.3 \\
(b) & 1.5 & 0.3 & 0.3 \\
(c) & 3 & 0 & 1 \\
(d) & 3 & 0.1 & 0.3 \\ \hline
\end{tabular}
\end{center}
\end{table}
Although the choice of the parameters of Table \ref{tab1} will
become clearer in Sec. \ref{sec4}, some comments can already be
made. This choice both depends on the structure of the sets $C_A$
and $C_B$ and on the characteristics of two particular extremals
denoted $X-$ and $Y-$ which start at the initial point, and
correspond respectively to a constant control equal to -1 and 1.
As detailed in the appendix \ref{appb}, the $X-$ and $Y-$
trajectories are either pseudo-periodic or aperiodic according to
the sign of the discriminant $\Delta=(\Gamma-\gamma_+)^2-4$ of the
system of Eqs. (\ref{eq5}). An exact resolution of the dynamics
shows that the trajectory of the system is aperiodic if $\Delta>0$
and pseudo-periodic otherwise. This point is summarized by the
diagram \ref{delta} of the appendix \ref{appb}. In Table
\ref{tab1}, we have chosen for three of the four examples $\Gamma$
such that $\Gamma>\gamma_++2$ to simplify the local structure of
the optimal synthesis around the fixed-point of the dynamics. A
pseudo-periodic trajectory is locally a spiral in the plane
$(x_2,x_3)$ around this fixed point which makes the analysis more
complex (see Sec. \ref{sec4}).
\subsection{Purity and limits of the dynamics} \label{sec3a}
Before analyzing the reachable set and the controllability of each
example, we begin by some general comments about the dynamics.

We first show that the field cannot locally compensate the effect
of dissipation. The purity of the quantum state is defined by the
function $2\textrm{Tr}[\rho^2]-1=x_2^2+x_3^2$. Pure states are
thus on the unit circle of the $(x_2,x_3)$ plane. Simple algebra
then leads to
\begin{equation} \label{eq15}
\frac{d(\textrm{Tr}[\rho^2])}{dt}=-\Gamma x_2^2+\gamma_-
x_3-\gamma_+ x_3^2 \ ,
\end{equation}
and we notice that this derivative does not depend on $u$, which
completes the proof. A more general proof of this point is given
in Refs. \cite{altafini1,altafini2}. We also point out that for
points $\textbf{x}$ where $\Delta_A(\textbf{x})<0$ then
$\frac{d\textrm{Tr}[\rho^2]}{dt}<0$ and inversely if
$\Delta_A(\textbf{x})>0$ then $\frac{d\textrm{Tr}[\rho^2]}{dt}>0$.
The curve $C_A$ divides the plane $(x_2,x_3)$ into a region where
the purity of the state locally increases and a region where it
locally decreases. On the boundary $C_A$, the purity is preserved.
This point can be qualitatively understood as follows. We recall
that the purity is equal to the square of the distance to the
origin. The conservative vector field $G$ is orthoradial (i.e.
normal to radial vectors) for each point $(x_2,x_3)\neq (0,0)$ of
the manifold. The dissipative vector field $F$ does not modify the
purity of the state if the radial component of $F$ vanishes i.e.
if $F$ is parallel to $G$ which is the definition of the curve
$C_A$.

We next analyze the fixed points of the dynamics when the field is
on, which are defined by $F+uG=0$. Since $F$ and $G$ are parallel,
the fixed points belong to the curve $C_A$. The field-free limit
point is the point of this line of maximum purity (see for
instance Fig. \ref{fig1}). This shows that the dissipation alone
allows to reach the state of maximum purity. Inversely, one can
ask if every point of the curve $C_A$ corresponds to a limit point
of the dynamics. The answer is positive for a real non-bounded
control $u$ since the limits can be written
\begin{eqnarray}\label{eq16}
\left\{ \begin{array}{ll}
x_2=\frac{-u\gamma_-}{\Gamma\gamma_++u^2} \\
x_3=\frac{\gamma_-}{\gamma_++u^2/\Gamma}
\end{array} \right. \ .
\end{eqnarray}
\subsection{Reachable sets} \label{sec3b}
We begin by recalling some results of Refs.
\cite{altafini1,altafini2} about controllability of dissipative
systems. A quantum dissipative system of finite dimension whose
dynamics is governed by the Lindblad equation is generically
accessible but not controllable. The accessibility property
characterizes the fact that the system can be driven in every
direction of the state space. Moreover, the concept of
accessibility does not take into account the reversibility or the
irreversibility of the process. The lack of controllability is
measured by the non small-time controllability of the system which
illustrates the irreversibility of the dynamics. This kind of
system is not small-time controllable because the field cannot
locally compensate the effect of dissipation as shown by Eq.
(\ref{eq15}). The accessibility property can be checked by the
computation of the dimension of the dynamical Lie algebra $L$ of
the system \cite{altafini1,altafini2}. We introduce for that
purpose the density matrix $\bar{\rho}$ of coordinates
$(1,x_2,x_3)$ which is given in the basis of the coherence vector
and we rewrite Eqs. (\ref{eq5}) in matrix form as follows
\begin{equation} \label{eq16a}
\dot{\bar{\rho}}=\mathcal{L}_F\bar{\rho}+u\mathcal{L}_G\bar{\rho}
\ ,
\end{equation}
where $\mathcal{L}_F$ and $\mathcal{L}_G$ are $3\times 3$ matrices
given by
\begin{eqnarray} \label{eq16b}
\mathcal{L}_F=\left(
\begin{array}{ccc}
0 & 0 & 0 \\ 0 & -\Gamma & 0 \\ \gamma_- & 0 & -\gamma_+
\end{array} \right) \textrm{and} \ \mathcal{L}_G=
\left(
\begin{array}{ccc}
0 & 0 & 0 \\ 0 & 0 & -1 \\ 0 & 1 & 0
\end{array} \right)
\ .
\end{eqnarray}
$L$ is the Lie algebra generated by $\mathcal{L}_F$ and
$\mathcal{L}_G$. If $\Gamma\neq \gamma_+$, a direct computation
shows that $L$ is either isomorphic to $\mathfrak{gl}(2)$ or to
the semi-direct sum $\mathfrak{gl}(2)\circledS \mathbb{R}^2$ of
respective dimensions 4 and 6 for $\gamma_-=0$ or $\gamma_-\neq
0$. The system is therefore accessible for
$\Gamma\neq \gamma_+$.\\

We now determine the reachable sets from their respective initial
states of the four examples. The reachable sets can be constructed
by the explicit construction of all the trajectories. We denote by
$\mathcal{R}(\textbf{x}_0)$ the reachable set from $\textbf{x}_0$.
 We first search for in this section the boundary of the reachable
sets. Then in Sec. \ref{sec4} we show that all the points inside
the boundary are attainable. We consider for that the $X-$ and
$Y-$ trajectories starting from the initial point and from the
field-free fixed point of the dynamics. A qualitative change
 occurs when the $X-$ or $Y-$ trajectories cross $C_A$ since the angle
between the vector $F(\textbf{x})$ and $G(\textbf{x})$ changes its
sign. The determination of all the extremal trajectories in Sec.
\ref{sec4} allows us to complete the construction.

\begin{figure}
\includegraphics[width=0.4\textwidth]{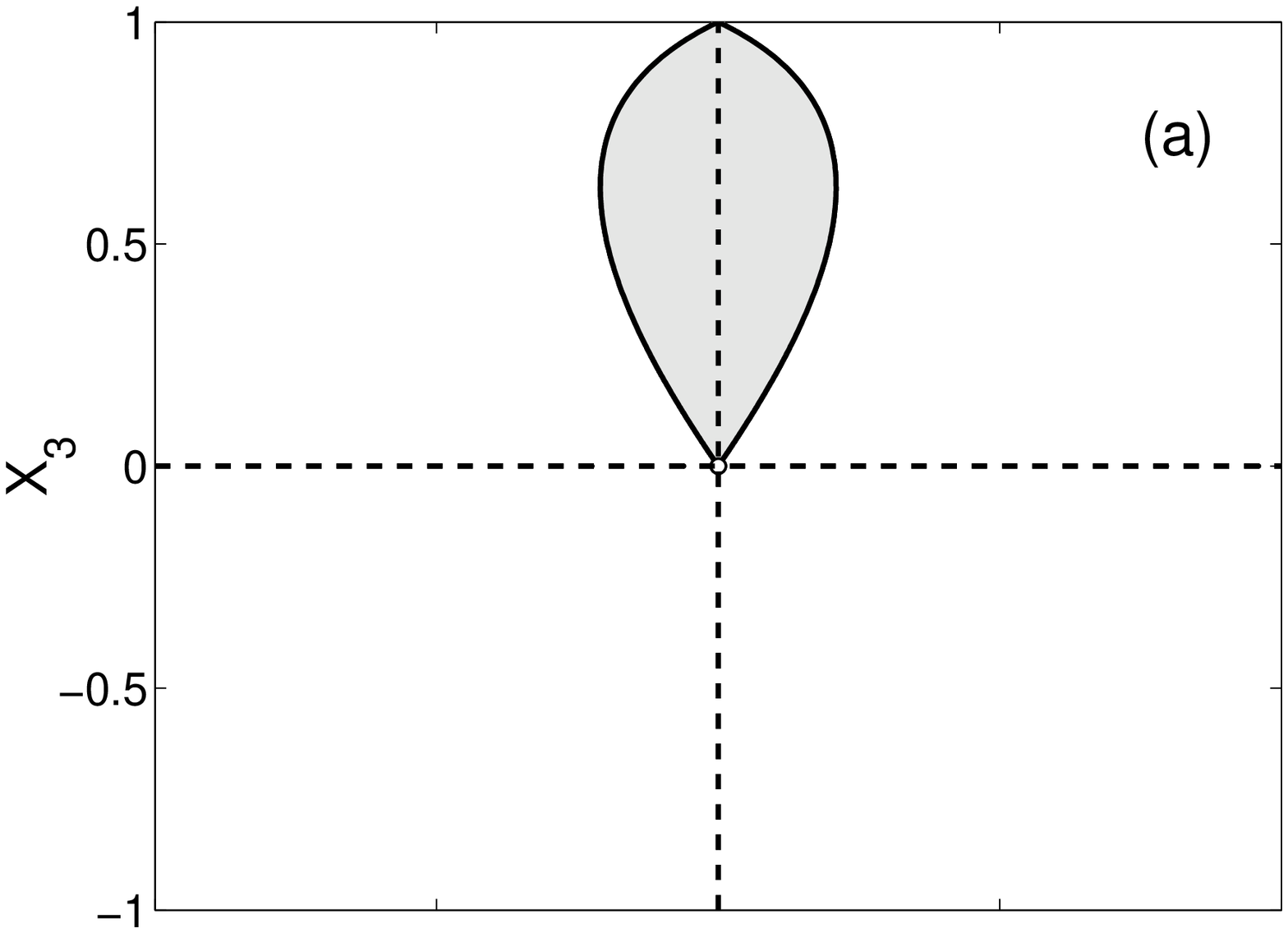}
\includegraphics[width=0.4\textwidth]{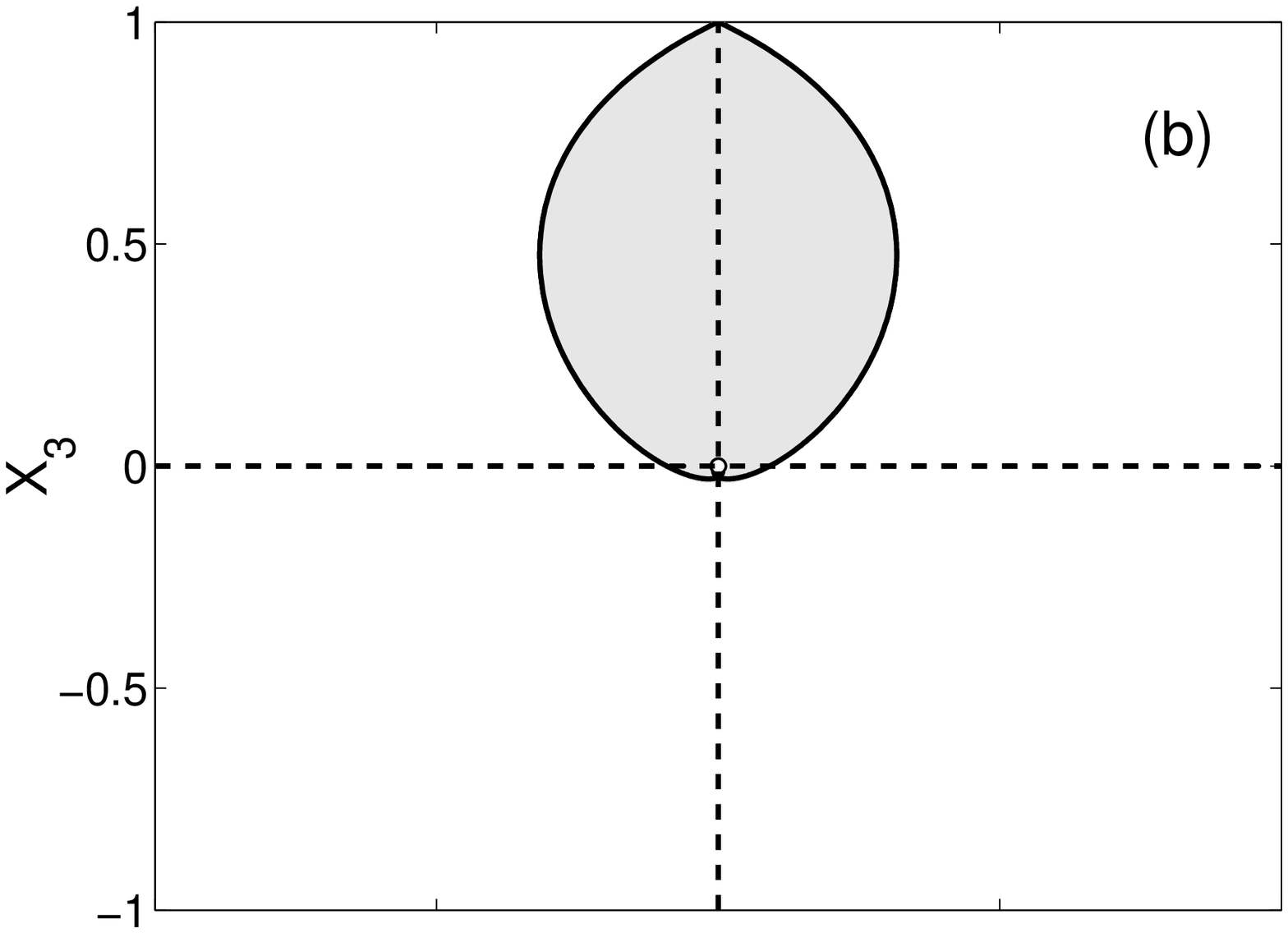}
\includegraphics[width=0.4\textwidth]{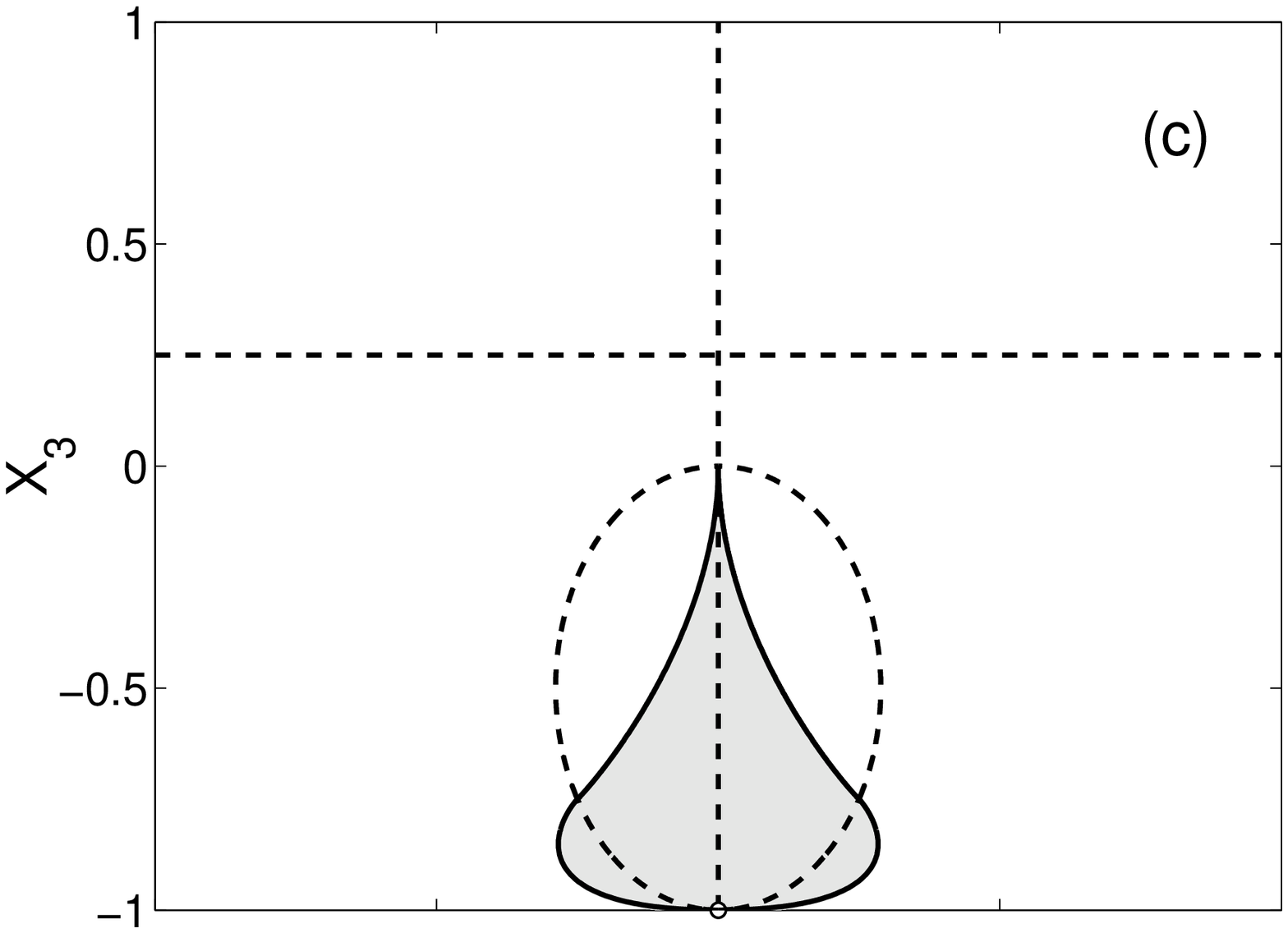}
\includegraphics[width=0.4\textwidth]{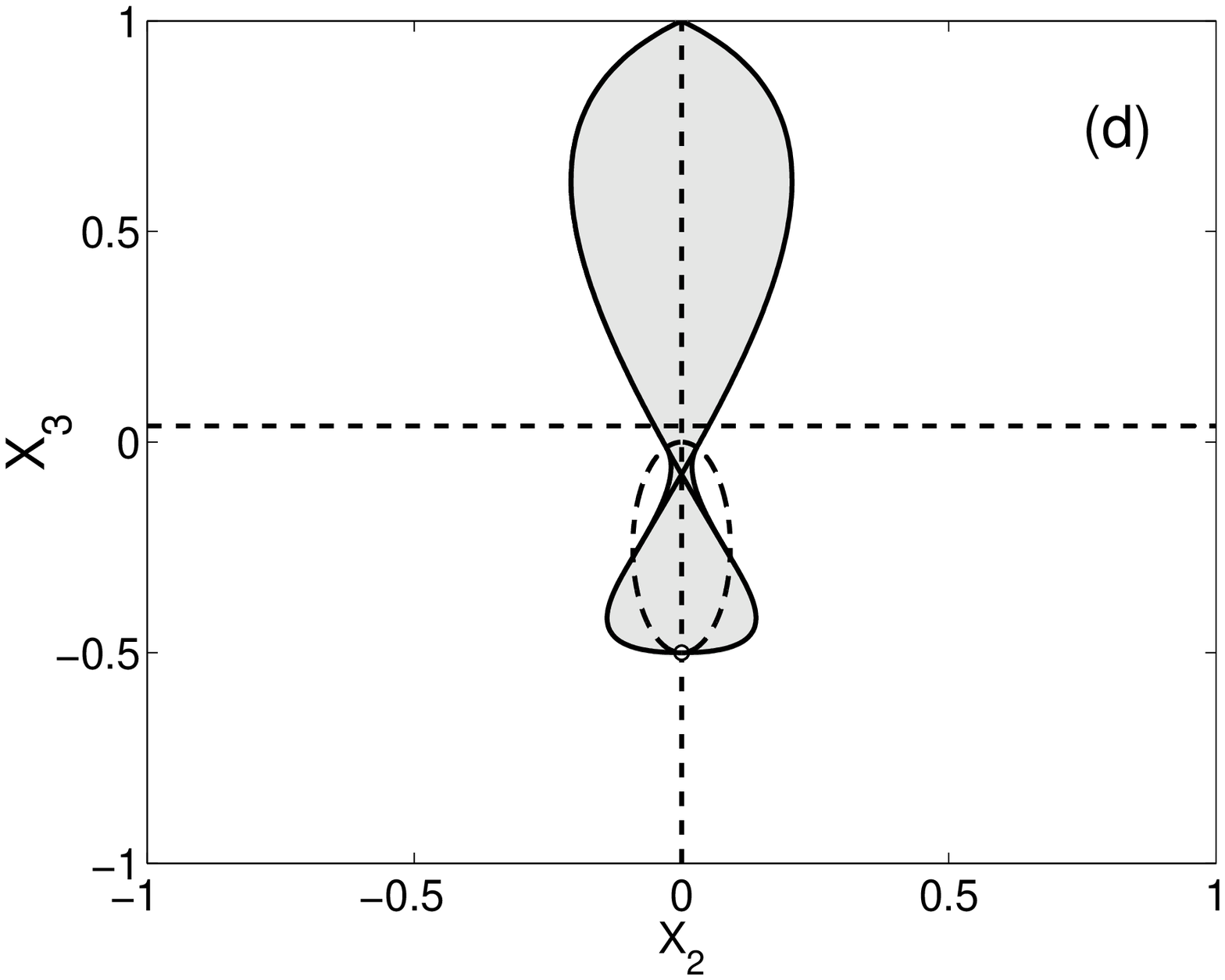}
\caption{\label{fig2} Reachable sets $\mathcal{R}$ (in grey) for
the cases (a), (b), (c) and (d). The sets $C_A$ and $C_B$ are
represented in dashed lines. The position of the field-free fixed
point is indicated by the small open circle. In (c) and (d), the
trajectories starting from the field-free fixed point do not
belong to $\mathcal{R}$ (see text).}
\end{figure}

We apply these remarks to the cases $(a)$ and $(b)$ where the
initial point has coordinates $(x_2=0,x_3=1)$ and $C_A$ is reduced
to the origin $(x_2=0,x_3=0)$. The origin is attained
asymptotically by the dynamics when $t\to +\infty$. The $Y-$ and
$X-$ trajectories, which have the same initial and final points,
are therefore global boundaries of the reachable set. This point
is illustrated in Figs. \ref{fig2}a and \ref{fig2}b for the cases
(a) and (b), where the reachable set $\mathcal{R}$ is in grey.

For the case $(c)$, the $X-$ and $Y-$ trajectories starting from
$(x_2=0,x_3=0)$ intersect asymptotically $C_A$. We also consider
the $X-$ and $Y-$ trajectories starting from the field-free fixed
point which define two new regions. As an infinite time is
necessary to reach this point, these two trajectories do not
belong to the reachable set. $\mathcal{R}$ is the union of all
these regions as shown in Fig. \ref{fig2}c. We will check in Sec.
\ref{sec4} by constructing the extremal trajectories that every
point of this set is effectively attainable.

For the case $(d)$ the situation is more difficult to analyze as
the $X-$ and $Y-$ trajectories starting from $(x_2=0,x_3=1)$
intersect transversally $C_A$. From this point of intersection,
the two trajectories do not correspond anymore to the boundary of
$\mathcal{R}$. After the crossing of $C_A$, the $X-$ or $Y-$
trajectories originating from the preceding curves define two new
regions of $\mathcal{R}$. We finally consider the limit point of
the field-free dynamics which can be attained as $t\to +\infty$
and the $X-$ and $Y-$ trajectories starting from this point. These
two lines define two new boundaries which intersect asymptotically
$C_A$ but which do not belong to $\mathcal{R}$. $\mathcal{R}$ is
therefore composed of all the regions constructed above and is
displayed in Fig. \ref{fig2}d. As before, only the determination
of the extremal trajectories will complete the construction.
\section{Time-optimal control} \label{sec4}
The application of the Pontryagin maximum principle is
particularly simple at least locally in the case of time minimum
control. The goal of this section is to solve this problem by
constructing the optimal synthesis i.e. the set of all the optimal
solutions starting form a given initial point $\textbf{x}_0$ and
reaching a point of $\mathcal{R}(\textbf{x}_0)$. For time
optimization, the optimal controls are composed of piecewise
constant parts ($u=\pm 1$) and of singular controls. Some
preliminary work in Sec. \ref{sec4a} has to be done before
determining the optimal solutions.
\subsection{Preliminary}\label{sec4a}
We define two functions $\Phi$ and $\theta$ \cite{boscain}.
$\Phi$, called the switching function, is given by
\begin{equation} \label{eq41}
\Phi(t)=\textbf{p}.G=-p_2x_3+p_3x_2 \ .
\end{equation}
Using Eq. (\ref{eq10}), standard considerations of maximization of
the pseudo-Hamiltonian $H_P$ lead to the conclusion that the
extremal field at time $t$ is given by $u=sign(\Phi(t))$ if
$\Phi(t)\neq 0$. A time $t_0$ such that the control changes sign
(i.e. such that $\Phi(t_0)=0$) is a switching time. If $\Phi$
vanishes on an interval $[t_0,t_1]$ then the corresponding
trajectory is called singular and is referred to as a $Z-$
trajectory. An important point is the fact that the $Z-$
trajectories lie in the set $C_B$ and that the corresponding
control $u=\phi$ is singular i.e. different from 1 or -1. $\phi$
can be calculated by imposing that
$\frac{d}{dt}\Delta_B(\textbf{x}(t))=0$ on $[t_0,t_1]$. In our
case, it can be shown that
\begin{equation} \label{eq41a}
\frac{d}{dt}\Delta_B(\textbf{x}(t))=0=\frac{\partial
\Delta_B}{\partial x_2}\dot{x_2}+\frac{\partial \Delta_B}{\partial
x_3}\dot{x_3} \ .
\end{equation}
One arrives after simple algebra to
\begin{equation} \label{eq41b}
\phi(\textbf{x})=\frac{-x_2\gamma_-(\Gamma-2\gamma_+)-2x_2x_3(\gamma_+^2-\Gamma^2)}
{2(\Gamma-\gamma_+)(x_2^2-x_3^2)-\gamma_-x_3} \ .
\end{equation}
For the line $x_2=0$ of $C_B$, this leads to $\phi(\textbf{x})=0$.
For the line $x_3=\frac{-\gamma_-}{2(\Gamma-\gamma_+)}$, we obtain
\begin{equation} \label{eq41c}
\phi(\textbf{x})=\frac{\gamma_-(\gamma_+-2\Gamma)}{2(\Gamma-\gamma_+)x_2}
\ .
\end{equation}
The control is admissible if $|\phi(\textbf{x})|\leq 1$ which
implies here the condition
\begin{equation} \label{eq41d}
|x_2|\geq |\frac{\gamma_-(\gamma_+-2\Gamma)}{2(\Gamma-\gamma_+)}|
\ .
\end{equation}
Not every $Z-$trajectory can be an optimal trajectory. We
introduce to characterize this point the notion of turnpike and
anti-turnpike curves \cite{boscain}. Let $\textbf{x}\notin C_A\cup
C_B$ and the function
$f(\textbf{x})=-\frac{\Delta_B(\textbf{x})}{\Delta_A(\textbf{x})}$.
A turnpike curve is an arc lying in $C_B$ such that for every
point $\textbf{x}$ of this arc $\Delta_A(\textbf{x})\neq 0$ and
$X(\textbf{x})$ and $Y(\textbf{x})$ are not tangent to $C_B$. It
is also assumed that $X(\textbf{x})$ and $Y(\textbf{x})$ point to
opposite sides of $C_B$ which define two regions $\Omega_x$ and
$\Omega_y$. If $f(\textbf{x})>0$ (resp. $f(\textbf{x})<0$) on
$\Omega_y$ and $f(\textbf{x})<0$ (resp. $f(\textbf{x})>0$) on
$\Omega_x$ then the arc is a turnpike (resp. anti-turnpike) arc.
The relation with the optimal synthesis can be stated as follows.
Using for instance the clock form $\alpha$ (see appendix
\ref{appa}), it can be shown that the anti-turnpike trajectories
are not optimal. Figure \ref{fig1} displays the turnpike and the
anti-turnpike curves for particular values of the parameters.
\\
The switching function $\Phi$ is a powerful tool that gives all
the informations on the extremal trajectory. However, a global
analysis of the extremals using only $\Phi$ is difficult because
all the initial values $(p_{20},p_{30})$ have to be tested. A more
global point of view is given by the function $\theta$ which has
the advantage not to depend on $\textbf{p}$.

Let $\textbf{v}$ be the vector of coordinates
$(\dot{x_2},\dot{x_3})$. Deriving Eqs. (\ref{eq5}) with respect to
time, it can be shown that $\textbf{v}$ satisfies the following
system of equations
\begin{eqnarray}\label{eq42}
\left\{ \begin{array}{ll}
\dot{v_2}=-\Gamma v_2-uv_3 \\
\dot{v_3}=-\gamma_+v_3+uv_2
\end{array} \right. \ .
\end{eqnarray}
We next introduce the vector $\tilde{v}(t)$. At time $t$, let us
assume that the dynamics reaches the point $(x_2(t),x_3(t))$ and
that the vector field $G$ at this point is given by
$G(x_2(t),x_3(t))=G(t)$. $\tilde{v}(t)$ is defined as the solution
at time 0 of Eqs. (\ref{eq42}) that at time $t$ is equal to
$G(t)$. The dynamics is thus propagated backwards during the time
$t$. $\theta$ is then defined as
\begin{equation} \label{eq43}
\theta(t)=\arg (\tilde{v}(0),\tilde{v}(t)) \ ,
\end{equation}
where the angle is measured counterclockwise. The definition of
$\theta$ originates from the fact that the product
$\textbf{p}(t)\cdot\textbf{v}(t)$ is a constant \cite{boscain}. We
thus have
\begin{equation} \label{eq43a}
\Phi(t)=p(t)\cdot G(t)=p(0)\cdot \tilde{v}(t) \ .
\end{equation}
The function $\theta$ allows to determine some properties of the
extremal trajectories. We briefly recall these points here. The
reader is referred to \cite{boscain} for rigorous definitions and
complete proofs. It can first be shown that
\begin{equation} \label{eq44}
sign(\dot{\theta})=sign(\Delta_B) \ ,
\end{equation}
which means that the zeros of $\dot{\theta}$ are located on $C_B$.
A switch can occur if $\dot{\theta}>0$ and $\theta>0$ or
$\dot{\theta}<0$ and $\theta<0$. In addition, the variation of
$\theta$ between two switches or between a switch and a singular
control is equal to 0 modulo $\pi$. This latter point can be
easily understood from Eq. (\ref{eq43a}). Other properties of
$\theta$ will be detailed during the construction of the optimal
synthesis.
\subsection{Time-optimal syntheses}\label{sec4b}
The parts \ref{synta}, \ref{syntb}, \ref{syntc} and \ref{syntd}
are rather technical and describe the way to obtain the optimal
syntheses. Section \ref{synte} details conclusions on the role of
the dissipation that can be gained from the resolution of the time
optimal control.
\subsubsection{Case (a)}\label{synta}
\begin{figure}
\includegraphics[width=0.4\textwidth]{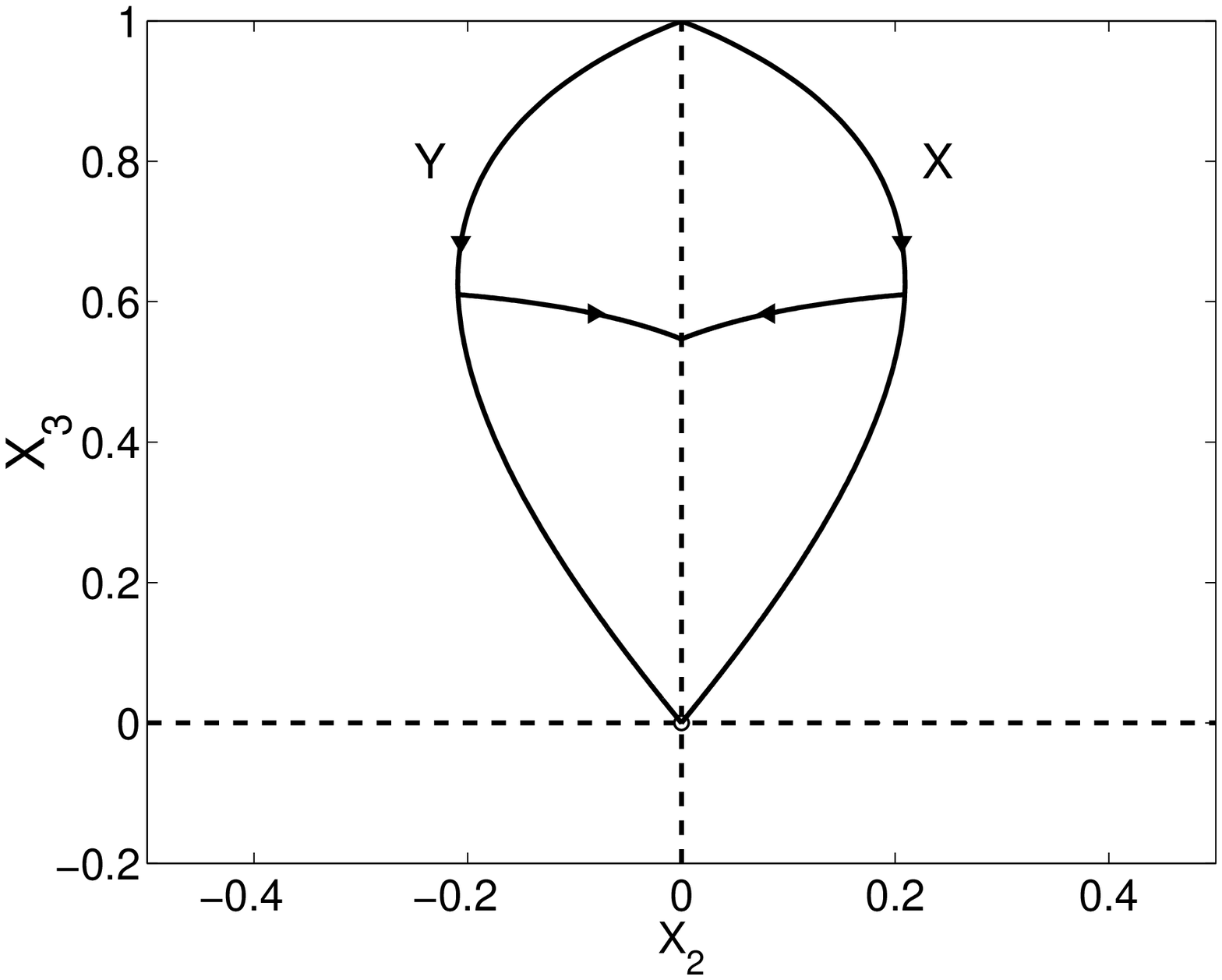}
\includegraphics[width=0.4\textwidth]{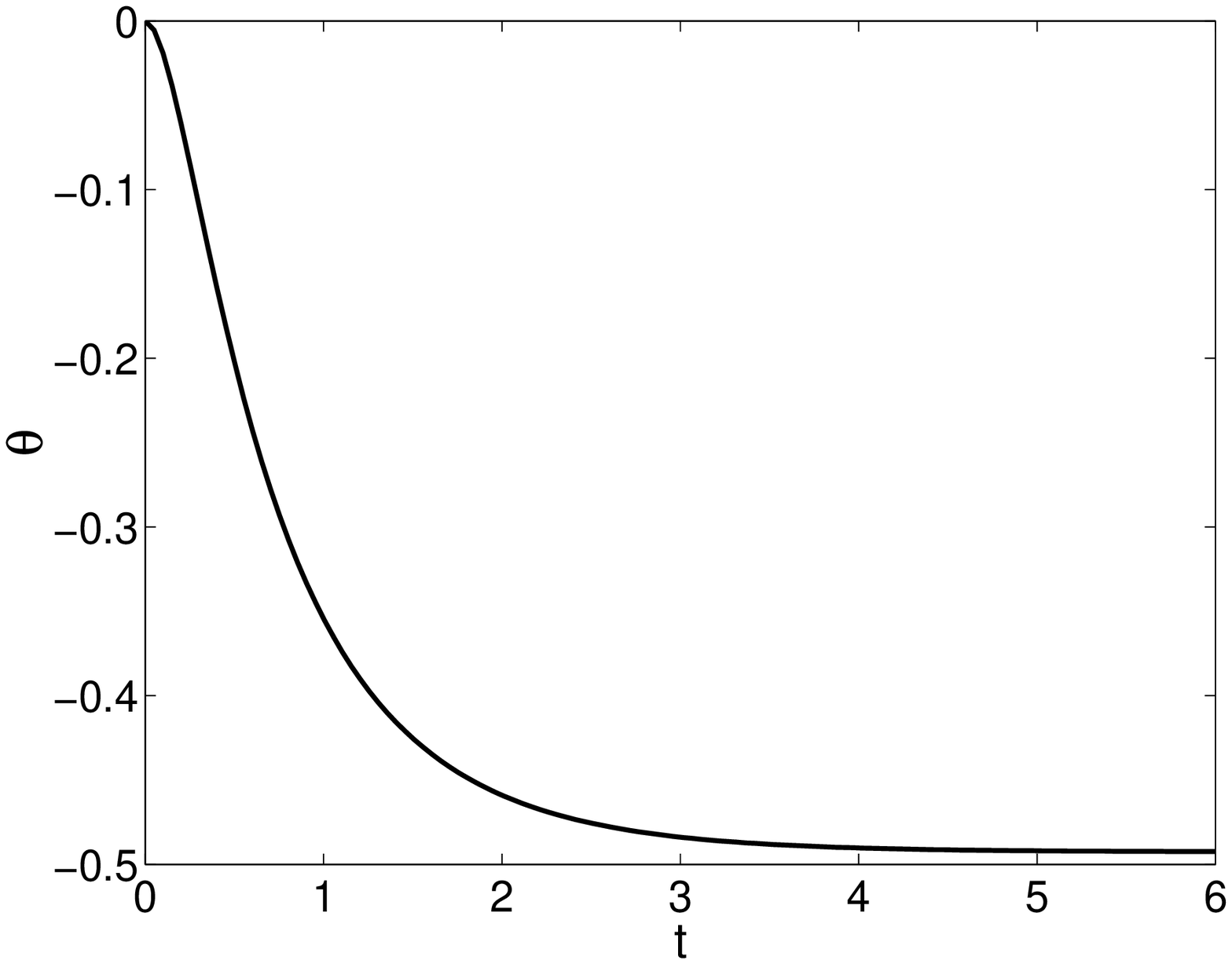}
\caption{\label{figsynta} The top figure represents the optimal
synthesis for the case (a) (see text). The dashed lines indicate
the locus $C_B$ and the small open circle the fixed point of the
dynamics. The bottom figure displays the evolution of $\theta$ as
a function of the time for the $Y-$ trajectory starting from the
point $(0,1)$.}
\end{figure}

In the case (a), the optimal trajectories are either bang or
bang-bang. A bang trajectory is a trajectory associated to a
single value of the control $u=+1$ or $u=-1$. A bang-bang
trajectory is the concatenation of an $X-$ and an $Y-$
trajectories. We denote by $X*Y$ such a concatenation where the
$Y-$ trajectory comes first. Here the maximum number of switches
is thus equal to 1. The elimination of extremal trajectories
selected by the Pontryagin maximum principle has been done through
the clock form $\alpha$ (see appendix \ref{appa} for details) and
the symmetry of the diagram with respect to the line $x_2=0$. We
recall that the clock form can only be used for trajectories
belonging to one of the four quadrants defined by $C_B$ and which
do not cross $C_A$. Figure \ref{figsynta} displays the optimal
synthesis for this problem and the evolution of the angle $\theta$
along the $Y-$ trajectory starting from the initial point $(0,1)$.
The plot of $\theta$ shows that the switches are always permitted
along the $X-$ or the $Y-$ trajectory. Note also that the form of
the curve representing $\theta$ is related to the sign of
$\Delta$. More precisely, if $\Gamma>\gamma_+ +2$ then $\theta$ is
a monotonically decreasing function and if $\Gamma<\gamma_+-2$
(case not treated here), $\theta$ starts increasing, passes
through a maximum when the trajectory crosses $C_B$ and then
decreases. The intermediate case $\gamma_+-2<\Gamma<\gamma_+ +2$
corresponds to the case (b). Using the clock form, it can be shown
that only one switch is possible for a trajectory in a given
quadrant. This point can also be determined from more general
considerations detailed in Ref. \cite{boscain}. The rest of the
optimal synthesis is deduced from the symmetry of the problem. The
line $x_2=0$ is called an overlap curve denoted by $K$ as it is
the locus of points reached by two optimal trajectories. We
finally notice that the dissipation alone is not used here by the
control and therefore cannot help accelerating the control. We
recall that the singular control on the vertical line of $C_B$ is
given by $u=0$.
\subsubsection{Case (b)}\label{syntb}
\begin{figure}
\includegraphics[width=0.4\textwidth]{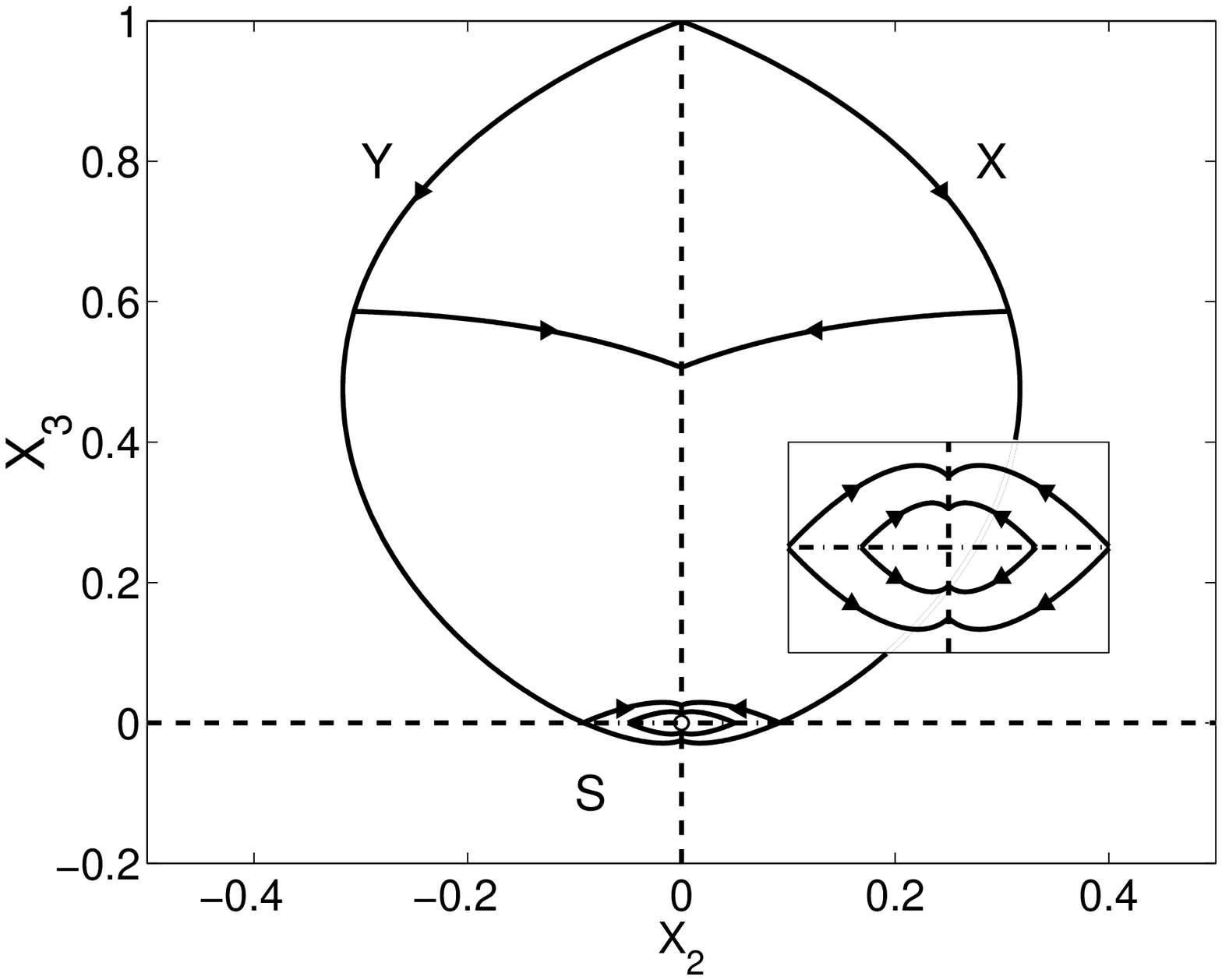}
\includegraphics[width=0.4\textwidth]{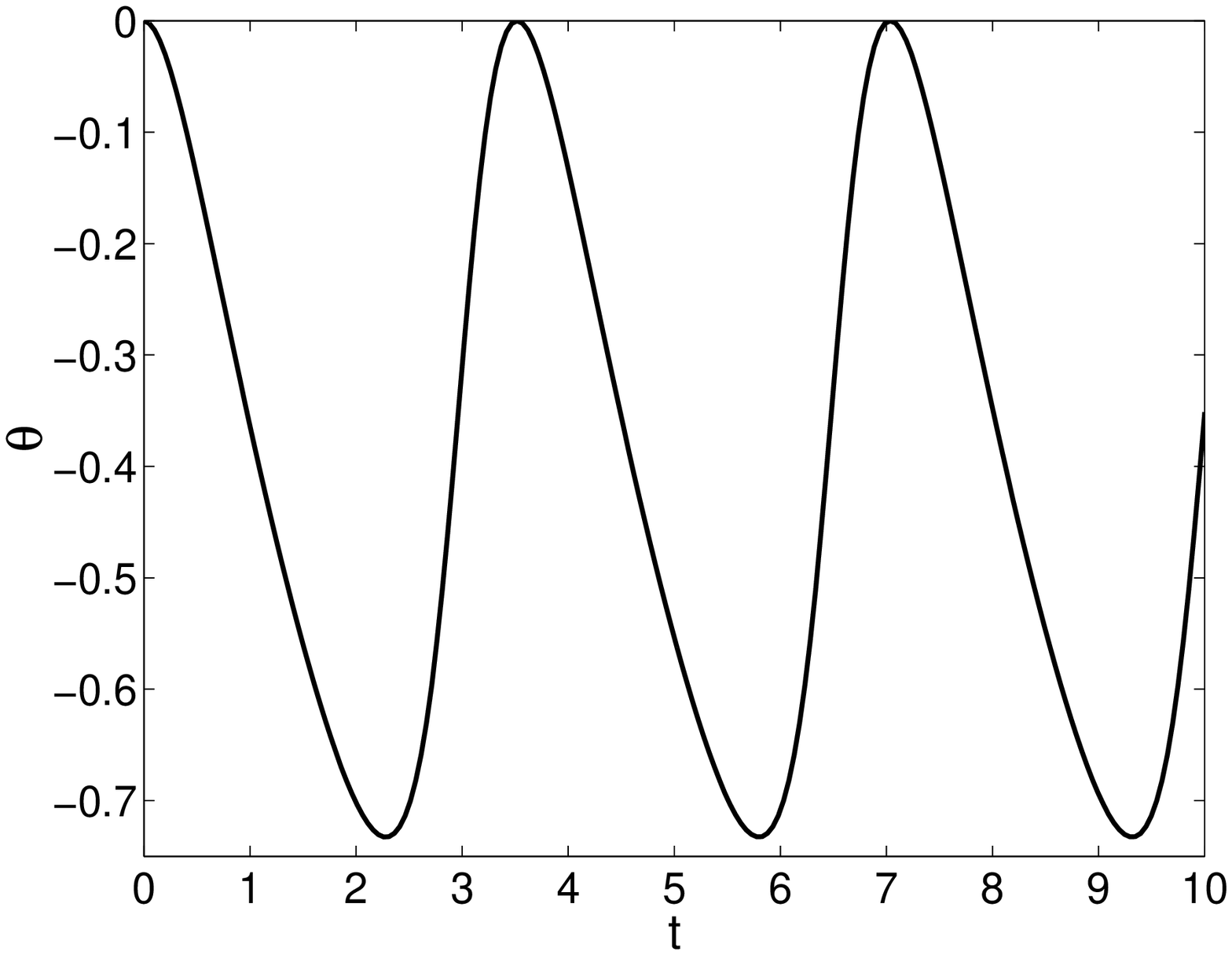}
\caption{\label{figsyntb} Same as Fig. \ref{figsynta} but for the
case (b). The dotted-dashed line represents the singular
trajectory $S$. The small insert is a zoom of the optimal
synthesis near the origin.}
\end{figure}

\begin{figure}
\includegraphics[width=0.4\textwidth]{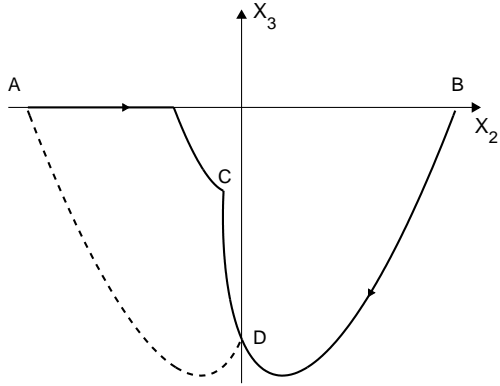}
\caption{\label{figschema} Use of symmetry to determine the
optimal trajectory (see text).}
\end{figure}

The situation is a little more complex in the case (b). For
$x_3>0$, the synthesis is similar to the one of case (a) i.e. the
trajectories are either bang or bang-bang. The function $\theta$
is here periodic, the times where $\dot{\theta}$ vanishes
correspond to the crossing of $C_B$. This function tells also us
that the initial $Y-$ and $X-$ trajectories cannot switch in their
respective second quadrant (i.e. the second quadrant they go
through) since $\dot{\theta}(t)>0$ and $\theta(t)<0$. By symmetry
with respect to the line $x_2=0$, one deduces that theses curves
are optimal up to this line. A singular control along the
horizontal line of $C_B$ is optimal from the point of intersection
between the initial $X-$ or $Y-$ trajectories and this line. Since
$\gamma_-=0$, the singular control is given by $\phi=0$. From this
singular line originate optimal trajectories with control $u=\pm
1$. Using the clock form $\alpha$, it can be shown that these
trajectories can not switch again. To summarize, in this second
part we have constructed optimal controls of the form $Y*Z*X$,
$Y*Z*Y$, $X*Z*X$ and $X*Z*Y$. This optimal synthesis is
represented in Fig. \ref{figsyntb}. Here, one sees that the
dissipation alone (with $u=0$) accelerates the control to reach a
point near the origin only along the horizontal direction and for
$x_3=0$. We finally give in Fig. \ref{figschema} an example of the
comparison of two extremal trajectories. The time to reach
respectively the points A and B from the initial point (0,1) by
the $Y-$ and $X-$ trajectories is the same. We would like
 to attain the point C from A or B. We use either the
$X-$trajectory from B or a concatenation of a $Y-$ and a $Z-$ ones
from A. The clock form $\alpha$ cannot be used here since the
trajectories belong to two different quadrants. We therefore
consider the symmetric image AD of the curve BD with respect to
$x_2=0$. The two extremals are thus in the same quadrant and
$\alpha$ can be used. We conclude that the curve $Y*Z*Y$ is
optimal to reach the point $C$.
\subsubsection{Case (c)}\label{syntc}
\begin{figure}
\includegraphics[width=0.4\textwidth]{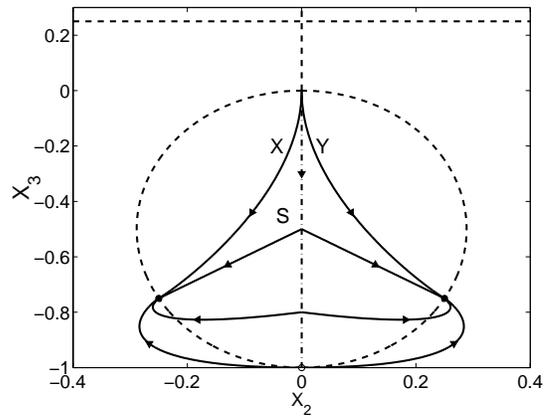}
\caption{\label{figsyntc} Same as Fig. \ref{figsyntb} but for the
case (c).}
\end{figure}

In this case, the singular line $x_2=0$ is optimal. The singular
control is equal to zero. Switches can occur from the initial $X-$
and $Y-$ trajectories but they do not lead to optimal
trajectories. Inversely, $X-$ and $Y-$ trajectories originating
from $S$ are found to be optimal. When two extremals cross $C_A$,
$\alpha$ cannot be used and a direct numerical comparison is then
performed. In this case, the function $\theta$ tells us that these
curves do not switch. Trajectories originating from $S$ that have
a switch are therefore not extremal. We finally point out that the
dissipation is beneficial for the control and decreases the
duration needed to purify the system. The optimal controls are
here of the form $X-$, $Y-$, $X*Z$ or $Y*Z$. Figure \ref{figsyntc}
displays the optimal synthesis for this problem.
\subsubsection{Case (d)}\label{syntd}
The case (d) is the more complex one and corresponds roughly to
the composition of cases (b) and (c). The difficulty lies in the
global structure of the control or, in other words, in gluing the
two preceding local analysis. For $x_3>0$, the optimal synthesis
is similar to the cases (a) or (b) with bang or bang-bang
trajectories. The bottom of the optimal synthesis from the point
of intersection of the initial $X-$ and $Y-$ trajectories is
similar to the case (c).

We now describe the central part of the synthesis. The horizontal
singular line of $C_B$ does not correspond to a singular
trajectory since $|\phi(\textbf{x})|>1$ which is a non-admissible
control. Consider the first point of intersection of $C_B$ with
the $X-$ or $Y-$ trajectory. Following \cite{boscain}, we know
that a switch curve $C$ originates from this point. To determine
the exact locus of $C$, we use the $\theta$ function since this
function varies by 0 modulo $\pi$ between two switches. $C$ is
constructed numerically. By definition of $C$, every trajectory
which crosses $C$ switches on $C$. We have found that $C$, $C_B$
and $C_A$ intersect in one point, the origin O.

Since the line $x_2=0$ is turnpike for
$\frac{\gamma_-}{\gamma_+}<x_3<0$, we can ask if this singular
trajectory is optimal i.e. if we can have a local optimal
synthesis of the form given by Fig. \ref{figlocal}. To answer this
question, we use the switching function $\Phi$. For
$\textbf{x}(t)\in C\cup S$, $\Phi(t)=0$ i.e. the vectors
$\textbf{p}(t)$ and $G(\textbf{x}(t))$ are orthogonal. Since the
direction of $G$ is known ($G$ is orthoradial), one can deduce the
direction of $\textbf{p}(t)$. Let $\textbf{z}_1$ and
$\textbf{z}_2$ be two points belonging respectively to $C$ and
$S$. The vectors $G(\textbf{z})$ associated to these points are
schematically represented in Fig. \ref{figlocal}. We let now the
states $\textbf{z}_1$ and $\textbf{z}_2$ go to $(0,0)$ and we
determine the directions of the different adjoint states. We
recall that the Pontryagin maximum principle states that
$\textbf{p}$ is a continuous function which does not vanish. When
$\textbf{z}_1$ goes to $(0,0)$, one deduces by a continuity
argument that $\textbf{p}_1$ is vertical in O. When $\textbf{z}_2$
goes to $(0,0)$, the limit direction of $\textbf{p}_2$ is given by
the switch curve $C$. To respect the continuity of $\textbf{p}$,
one sees that $C$ has to be tangent to the line $x_2=0$ in O. Due
to the complexity of analytical calculations, we have checked
numerically that this is not the case. The singular line for
$x_3<0$ is therefore not optimal. The trajectories of the form
$Y*X*Y$ or $X*Y*X$ are thus optimal up to $x_2=0$. In addition,
when the initial $X-$ and $Y-$ trajectories cross $C_A$, the angle
between the vectors $F(\textbf{x})$ and $G(\textbf{x})$ changes
its sign. New optimal trajectories originate from this point of
intersection and correspond to two new regions of the reachable
set.
\begin{figure}
\includegraphics[width=0.4\textwidth]{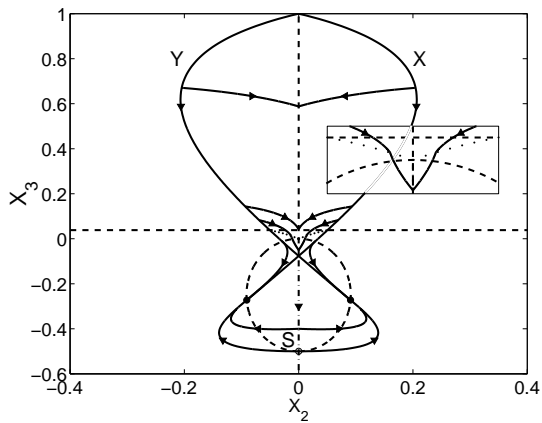}
\caption{\label{figsyntd} Same as Fig. \ref{figsyntb} but for the
case (d). The dotted line represents the switch curve $C$.}
\end{figure}
\begin{figure}
\includegraphics[width=0.4\textwidth]{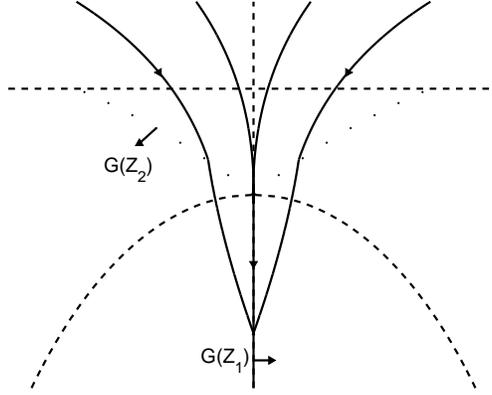}
\caption{\label{figlocal} Possible optimal synthesis around the
origin.}
\end{figure}
\subsubsection{Qualitative conclusions on the
dynamics}\label{synte}

From the results obtained in the preceding sections, some
qualitative conclusions can be made with respect to the
dissipation effect on the time optimal control of the dynamics.
The dissipation is not undesirable when the dissipation allows to
purify the system [cases (c) and (d)] and help accelerating the
control. In contrast for the conversion of a pure state into a
mixed state, the dissipation alone increases the duration of the
control and its effect is not beneficial for the control [cases
(a) and (b)]. The example (d) summarizes well the situation. As
long as the purity of the state decreases, it is advantageous to
use a control field but when the purity starts increasing the
dissipation alone becomes more efficient.

Finally, we point out that a visual inspection of the optimal
syntheses allows to check the form of the reachable sets given in
Sec. \ref{sec3b}.
\section{Conclusion} \label{sec6}
We have investigated the time-optimal control of a two-level
dissipative system. For different initial and final points, we
have constructed the optimal syntheses which allow us to
rigorously conclude on the beneficial role of the dissipation in
order to reduce the duration of the process. An open question is
the generalization of the present study to more complex quantum
systems with for instance two control parameters or three levels.
The determination of the optimal syntheses for these systems
requires more sophisticated mathematical tools than those used in
this paper \cite{bonnard}. In addition, such an analysis could
also be performed on other problems of quantum control. We point
out that the question of measurement in the context of geometrical
control theory is particularly interesting. Measurements can be
viewed as a non-unitary control which allows to enlarge the
reachable sets. For instance, we can consider systems not
completely controllable by unitary control but which become
controllable if measurements are used. A simple example is given
by a finite number of levels of the harmonic oscillator with a
constant dipolar interaction only between adjacent energy levels
\cite{fu}. From a formal point of view, this problem raises the
challenge of the precise formulation of the Pontryagin maximum
principle in this context which is to our knowledge an open
question. We notice that all these geometrical tools would allow
us to answer some physical relevant questions such as the time at
which measurements have to be done to minimize the duration of the
control.
\appendix
\section{The clock form} \label{appa}
\begin{figure}
\includegraphics[width=0.4\textwidth]{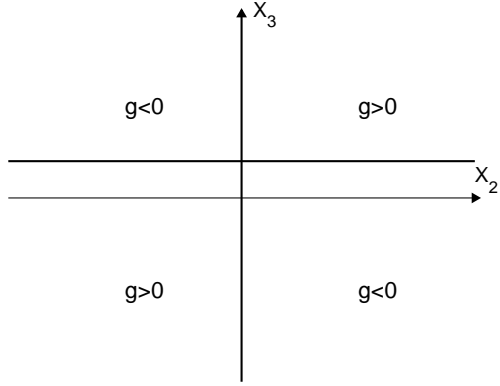}
\caption{\label{figf} Sign of the function $g$ in the plane
$(x_2,x_3)$. The large solid lines indicate the position of the
set $C_B$.}
\end{figure}

We derive in this section the expression of the clock form denoted
$\alpha$ \cite{bonnard}. By definition, the clock form is a 1-form
which fulfills the following conditions
\begin{eqnarray}\label{eqapp1}
\left\{ \begin{array}{ll}
\alpha(F)=1 \\
\alpha(G)=0
\end{array} \right. \ .
\end{eqnarray}
A solution of this system exists except on the set $C_A$ where $F$
and $G$ are collinear. If we write $\alpha$ as
$\alpha=\alpha_2dx_2+\alpha_3dx_3$ then simple algebra shows that
$\alpha_2$ and $\alpha_3$ are solutions of the system
\begin{eqnarray}\label{eqapp2}
\left\{ \begin{array}{ll}
\alpha_2(-\Gamma x_2)+\alpha_3(\gamma_--\gamma_+x_3)=1 \\
\alpha_2x_3=\alpha_3x_2
\end{array} \right. \ .
\end{eqnarray}
We obtain that
\begin{eqnarray}\label{eqapp3}
\left\{ \begin{array}{ll}
\alpha_2=\frac{-x_2}{\Gamma x_2^2-\gamma_-x_3+\gamma_+ x_3^2} \\
\alpha_3=\frac{-x_3}{\Gamma x_2^2-\gamma_-x_3+\gamma_+ x_3^2}
\end{array} \right. \ .
\end{eqnarray}
From the 1-form $\alpha$, we can define the 2-form $d\alpha$ which
is given by
\begin{equation} \label{eqapp4}
d\alpha=(\frac{\partial \alpha_3}{\partial x_2}-\frac{\partial
\alpha_2}{\partial x_3})dx_2\wedge dx_3 \ ,
\end{equation}
and reads after some calculations
\begin{equation} \label{eqapp5}
d\alpha=\frac{2\Gamma x_2 x_3+\gamma_-x_2-2\gamma_+
x_2x_3}{[\Gamma x_2^2-\gamma_-x_3+\gamma_+x_3^2]^2} dx_2\wedge
dx_3 \ .
\end{equation}
If we write $d\alpha$ as $d\alpha=g(x_2,x_3)dx_2\wedge dx_3$ then
one sees that $g(x_2,x_3)=0$ on $C_B$ and that the function $g$
has a constant sign in the regions delimited by the lines of
$C_B$. This point is displayed in Fig. \ref{figf}.
As suggested by its name, the clock form is a form which allows
one to determine the time taken to travel a path and to compare
the extremals. Let $\gamma$ be a path in the plane $(x_2,x_3)$
which does not cross $C_A$ and $T$ the time of travel along
$\gamma$. We have
\begin{equation} \label{eqapp5a}
\int_\gamma\alpha=\int_0^T\alpha(\dot{\textbf{x}})dt=\int_0^T
\alpha(F)dt=T \ .
\end{equation}
We consider now two paths $\gamma_1$ and $\gamma_2$ starting and
ending at the same points and respectively associated to the
durations $T_1$ and $T_2$. One shows by using the Stokes theorem
that
\begin{equation} \label{eqapp5b}
T_1-T_2=\int_{\gamma_1}\alpha-\int_{\gamma_2}\alpha=\int_D d\alpha
\ ,
\end{equation}
where $D$ is the surface delimited by $\gamma_1\cup -\gamma_2$.
For paths $\gamma_1$ and $\gamma_2$ which lie in one of the four
quadrants defined by $C_B$, we can straightforwardly determine the
time-optimal trajectory.
\section{Analytical determination of the dynamics} \label{appb}
\begin{figure}
\includegraphics[width=0.4\textwidth]{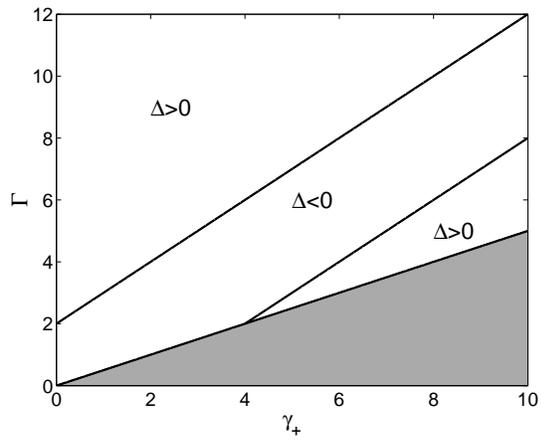}
\caption{\label{delta} Sign of the discriminant $\Delta$ as a
function of the parameters $\Gamma$ and $\gamma_+$. The parameters
are in arbitrary units. The zone in grey is excluded for Lindblad
dynamics.}
\end{figure}

In this section, we solve analytically the dynamics of the system
given by the following system of differential equations
\begin{eqnarray}\label{eqapp6}
\left\{ \begin{array}{ll}
\dot{x_2}=-\Gamma x_2-ux_3 \\
\dot{x_3}=\gamma_--\gamma_+x_3+ux_2
\end{array} \right. \ .
\end{eqnarray}
We assume that $u=\pm 1$ and we denote by $u=\varepsilon$ the
control term. Combining Eqs. (\ref{eqapp6}), one arrives to
equations that depend on only one variable $x_2$ or $x_3$
\begin{eqnarray}\label{eqapp7}
\left\{ \begin{array}{ll}
\ddot{x_2}+(\Gamma+\gamma_+)\dot{x_2}+(1+\Gamma\gamma_+)x_2+\varepsilon \gamma_-=0 \\
\ddot{x_3}+(\Gamma+\gamma_+)\dot{x_3}+(1+\Gamma\gamma_+)x_3-\Gamma\gamma_-=0
\end{array} \right. \ .
\end{eqnarray}
Equations (\ref{eqapp7}) are linear inhomogeneous differential
equations of second order whose discriminant $\Delta$ is given by
\begin{equation} \label{eqapp8}
\Delta=(\Gamma-\gamma_+)^2-4 \ .
\end{equation}
A qualitative change of the solutions is expected according to the
sign of $\Delta$, i.e. a transition from a pseudo-periodic
solution for $\Delta<0$ to an aperiodic solution for $\Delta\geq
0$. This point is summarized in Fig. \ref{delta}. Note that the
parameter $\gamma_-$ plays no role in the structure of these
trajectories but modifies their limits points. Simple algebra
allows one to calculate the exact solutions. For instance for
$\Delta>0$, we obtain
\begin{eqnarray}\label{eqapp8}
\left\{ \begin{array}{ll}
x_2(t)=e^{-(\Gamma+\gamma_+)t/2}[x_{20}\cosh(\frac{\delta}{2}t)-(x_{20}\frac{\Gamma}{\delta}
-x_{20}\frac{\gamma_+}{\delta}+\frac{2\varepsilon}{\delta}x_{30})\sinh(\frac{\delta}{2}t)]
\\
x_3(t)=e^{-(\Gamma+\gamma_+)t/2}[x_{30}\cosh(\frac{\delta}{2}t)+
(\frac{(\Gamma-\gamma_+)}{\delta}x_{30}+\frac{2\varepsilon}{\delta}x_{20}+\frac{2\gamma_-}{\delta})\sinh(\frac{\delta}{2}t)]
\end{array} \right. \ .
\end{eqnarray}
where $(x_{20},x_{30})$ is the initial point of the dynamics and
$\delta=\sqrt{\Delta}$.
\begin{acknowledgments}
The authors thank B. Bonnard and U. Boscain for many helpful
discussions.
\end{acknowledgments}

\end{document}